\documentclass[aps,prl,showpacs,eqsecnum,twocolumn,superscriptaddress]{revtex4}
\usepackage{amsmath,amssymb}
\usepackage{graphicx}

\begin{document}

\title{Exploring binary-neutron-star-merger scenario 
of short-gamma-ray bursts by gravitational-wave observation}

\author{Kenta Kiuchi}
\affiliation{Department of Physics, Waseda University, 3-4-1
 Okubo, Shinjuku-ku, Tokyo 169-8555, Japan~}

\author{Yuichiro Sekiguchi}
\affiliation{Division of Theoretical Astronomy/Center for
 Computational Astrophysics, National Astronomical Observatory of
 Japan, 2-21-1, Osawa, Mitaka, Tokyo, 181-8588, Japan~}

\author{Masaru Shibata}
\affiliation{Yukawa Institute for Theoretical Physics, 
Kyoto University, Kyoto, 606-8502, Japan~} 

\author{Keisuke Taniguchi} \affiliation{Department of Physics,
 University of Wisconsin-Milwaukee, P.O. Box 413, Milwaukee, Wisconsin 53201}

\date{\today}

\begin{abstract}
We elucidate the feature of gravitational waves (GWs) from binary
neutron star merger collapsing to a black hole by general relativistic
simulation. We show that GW spectrum imprints the coalescence
dynamics, formation process of disk, equation of state for neutron
stars, total masses, and mass ratio. A formation mechanism
of the central engine of short $\gamma$-ray bursts, which are likely
to be composed of a black hole and surrounding disk, therefore could be
constrained by GW observation. 
\end{abstract}


\maketitle

\emph{Introduction.}---Coalescence of binary neutron stars (BNSs) is
one of the most promising sources for kilometer-size laser 
interferometric gravitational-wave detectors 
~\cite{LIGO}.
It is also a candidate for the central engine of short $\gamma$-ray
bursts (GRBs), which emit huge energy $\agt 10^{48}$ ergs in a short
time scale $\sim 0.1$--1 s~\cite{Narayan:1992iy,Zhang:2003uk}.
According to a standard scenario of GRBs based on the so-called merger
scenario, a stellar-mass black hole (BH) surrounded by a hot and
massive disk (or torus) should be formed after the merger. Possible
relevant processes to extract the energy of this BH-accretion disk
system for launching a relativistic jet are neutrino-anti neutrino
($\nu\bar{\nu}$) annihilation \cite{Piran:2004ba} and/or magnetically
driven mechanisms, so-called Blandford-Znajek process~\cite{BZ}.
Studies based on the $\nu\bar{\nu}$ annihilation scenario suggest that
an accretion rate of $\dot{M}\agt 0.1M_{\odot}$/s is required to
achieve a sufficiently high energy efficiency~\cite{GRBdiskA} and that
if the disk had a mass $\agt 0.01M_\odot$, it could supply the
required energy by neutrino radiation for duration of $\agt 100$ ms
~\cite{GRBdiskN}.  General relativistic magnetohydrodynamic
simulations also indicate that if a rapidly rotating BH is formed, its
rotational energy can be extracted by the Blandford-Znajek process to
achieve a high energy efficiency~\cite{GRBMHD}.  However, the merger
scenario has not been proven yet observationally, because the counter
part for most of short GRB has not been identified
\cite{Fox:2005kv}. Gravitational waves (GWs) are much more transparent
than electromagnetic waves with respect to scattering with matter, and
hence, they can propagate from extremely dense region with negligible
scattering.  Therefore, GWs can be valuable observable for determining
the central engine of GRBs \cite{KM} and for exploring its formation
process. This fact motivates to study coalescence of BNSs and emitted
GWs.

For theoretically studying the late inspiral, merger, and ringdown
phases of BNSs, numerical relativity is the unique
approach. Simulations of BNS merger
~\cite{Shibata:2002jb,Baiotti:2008ra,Kiuchi:2009jt} suggest that (i)
if the total mass of the BNS is smaller than a threshold mass ($M_{\rm
thr}$) a hypermassive neutron star (HMNS) sustained by rapid and
differential rotation is formed after the merger and will survive for
more than $\sim 100$ ms~\cite{Shibata:2002jb}, and (ii) otherwise, a
BH is formed after the merger in a dynamical time scale $\sim 1$ ms.
In the former case, GW emission \cite{Shibata:2002jb} or
magnetic-field effect such as the magnetorotational instability
~\cite{DLSSS} or neutrino radiation \cite{janka} will play an
essential role in the subsequent evolution of the HMNS, which will
eventually collapse to a BH.  Because of the complexity, it will not
be an easy task to theoretically clarify the formation process of the
BH and gravitational waveforms as well as to accurately predict the
launching mechanism of short GRBs.  By contrast, in the later case,
the formation process of a BH and surrounding disk is primarily
determined by the general relativistic hydrodynamics, and magnetic
field or neutrino radiation will play a minor role.  Reliable
theoretical prediction of merger process and gravitational waveforms
by numerical relativity is feasible in this case.  Motivated by this
idea, we systematically performed numerical relativity simulations to
clarify the relation between possible formation process of GRB's
central engine (a system composed of a BH and surrounding disk) and
gravitational waveforms, and suggest that GW observation could 
constrain the formation process for the central engine of short GRBs.

\emph{Method and initial models.}--- Formulation and numerical schemes
for solving Einstein's equations, hydrodynamics, and other techniques
such as GW extraction are essentially the same as those in
Ref.~\cite{Kiuchi:2009jt}, to which the reader may refer for details.
For modeling the EOS of neutron stars, we adopt the
Akmal-Pandharipande-Ravenhall (APR)~\cite{APR:1998}, Skyrme-Lyon
(SLy)~\cite{Douchin:2001}, and Friedman-Pandharipande-Skyrme
(FPS)~\cite{Friedman:1981} EOSs for zero-temperature part.
Shock-heating effect, although it gives only minor contribution, is
mimicked by $\Gamma$-law EOS with $\Gamma=2$.  BNSs of the
irrotational velocity field and zero-temperature in quasiequilibrium
circular orbits are prepared as initial conditions, which are computed
by the LORENE library codes~\cite{Gourgoulhon:2000nn}.  For all the
models, the initial orbital separation is large enough that the BNSs
spend about 4 orbits in the inspiral phase before the merger.  Grid
structure and resolution are approximately the same as in
Ref.~\cite{Kiuchi:2009jt}; the major diameter of the neutron stars is
covered by about 80 uniform grids and outer boundaries of numerical
domain are located at about 1.2$\lambda_0$ where $\lambda_0$ is
initial gravitational wavelength.  Total mass, $m_0$, is chosen to be
large enough that a BH is formed in a dynamical time scale ($\sim 1$
ms) after the onset of the merger.  This can be achieved for $m_0 >
M_{\rm thr},$ where $M_{\rm thr}\approx 2.8$--$2.9M_\odot$ for the
APR, $\approx 2.7$--$2.8M_\odot$ for the SLy, and $\approx
2.5$--$2.6M_\odot$ for the FPS EOS,
respectively~\cite{Shibata:2002jb}.  Total mass and mass ratio are
chosen in the range of $2.9M_\odot\le m_0 \le 3.1M_\odot,0.8\le \nu
\equiv m_1/m_2 \le 1$ for the APR, $2.8M_\odot\le m_0 \le 3.0
M_\odot,\nu=1$ for the SLy, and $2.6M_\odot\le m_0 \le 2.8
M_\odot,\nu=0.8$ and $1$ for the FPS EOS.  $m_{1}$ and $m_2$ $(m_2\ge
m_1)$ denote the gravitational mass of two neutron stars in isolation
and $m_0=m_1 + m_2$.  We name the models according to the EOS, $m_0$,
and $\nu$; e.g., A2.9-0.9 is modeled by the APR EOS with
$m_0=2.9M_\odot$ and $\nu=0.9$.

\begin{figure}
\centerline{
\includegraphics[width=9.0cm]{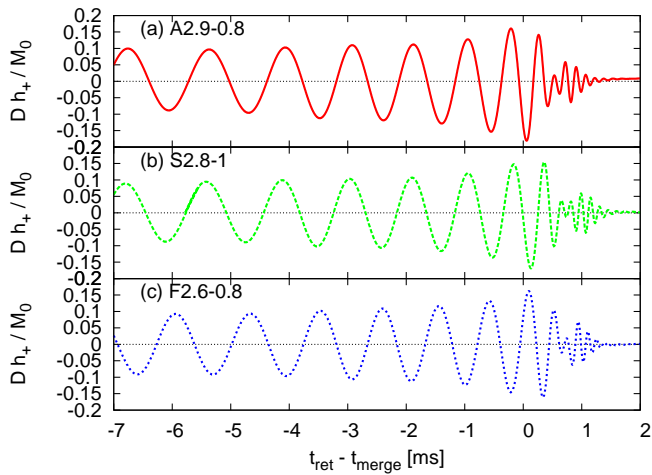}
}
\vspace{-3mm}
\caption{\label{fig:fig1} $+$ modes of GWs for
  (a) A2.9-0.8, (b) S2.8-1, and (c) F2.6-0.8. 
  $D$ is the distance from the source to the observer, who is
  located along the axis perpendicular to the orbital plane.  
$t_{\rm merge}$ denotes approximate time at the onset of the merger.}
\end{figure}

\emph{Results.}---We investigate the relation between GWs 
and possible formation process of a BH-disk system in a wide range 
of parameter space. First of all, we briefly review a typical 
coalescence process of BNSs collapsing dynamically to a BH 
(see Ref.~\cite{Kiuchi:2009jt} in details).
During the inspiral phase, the BNSs adiabatically evolve gradually
decreasing its orbital separation due to the gravitational radiation
reaction.  GWs in this phase are characterized by a chirp signal (see
the waveform for $t_{\rm ret}-t_{\rm merge} \alt 0$ in
Fig.~\ref{fig:fig1}).  After the binary separation becomes $\alt 3$
neutron star radii, the merger sets in.  For $\nu \approx 1$, a
high-density region with density $> 10^{15}~{\rm g/cm^3}$ is
subsequently formed at the center, and then, collapses to a BH within
$\sim 1$ ms.  During this merger process, small spiral arms, composed
of two symmetric arms, are formed around the BH but they are
eventually swallowed by the BH.  The final outcome is a rotating BH
nearly in a vacuum state.  For $\nu \alt 0.9$, the less massive star
is tidally deformed in a close orbit and disrupted just after the
onset of the merger. The outer part of the tidally disrupted
neutron-star matter forms an asymmetric spiral arm which is more
massive and widely-spread than that for the equal-mass case.  Although
the central high-density region, formed at the onset of the merger,
collapses to a BH, the large spiral arm subsequently forms an
accretion disk around the BH because it has angular momentum large
enough to escape from the capture by the BH. GWs in the merger phase
are characterized by the short-term burst-type waves $(0 \alt t_{\rm
ret}-t_{\rm merge} \alt 1$ ms in Fig.~\ref{fig:fig1}) and after the BH
formation, quasinormal modes (QNM) are excited ($t_{\rm ret} - t_{\rm
merge} \agt 1~{\rm ms}$ in
Fig.~\ref{fig:fig1}). Figure~\ref{fig:fig1}, which plots GWs for
A2.9-0.8, S2.8-1, and F2.6-0.8 and shows typical waveforms for the BNS
merger, illustrates features mentioned above.

\begin{figure}
\centerline{
\includegraphics[width=7.8cm]{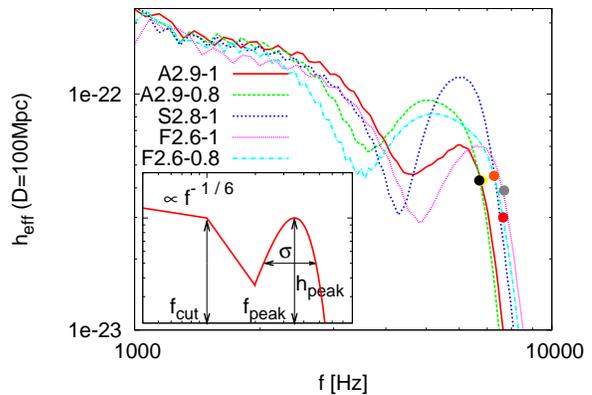}
}
\caption{\label{fig:fig2} Effective amplitude for models A2.9-1,
   A2.9-0.8, S2.8-1, F2.6-1, and F2.6-0.8. The filled circles denote 
  the QNM frequency of the formed BHs. The small panel shows a
  schematic figure of the GW spectrum.  }
\end{figure}

GW spectrum more clearly reflects dynamics of the merger process.
Figure~\ref{fig:fig2} plots an effective amplitude of GWs as a
function of the frequency, $h_{\rm eff}(f)$, for models A2.9-0.8,
A2.9-1, S2.8-1, F2.6-1, and F2.6-0.8. The effective amplitude is
defined by $h_{\rm eff}\equiv \tilde{h}(f) f m_0/D $ with
$\tilde{h}(f)$ being the Fourier transform of $h_+ +ih_{\times}$. We
assume that the distance from the observer to the source is $D=100$
Mpc, because more than one event per year is predicted for such a
large distance \cite{Kalogera}.  Note that GW frequency increases as
the orbital separation decreases and as the compactness of the merged
object increases: GWs with $f \alt 1$ kHz, $1 \alt f \alt 3$ kHz, and
$f \agt 3$ kHz are emitted in the inspiral phase, early merger, and
late merger phases, respectively.

The effective amplitude gradually decreases for $f\alt 3$ kHz
according to $\propto f^{-n}$ with $n\approx 1/6$ which is the
typical number for the inspiral phase, and for $f \agt f_{\rm cut}
\approx 2.5$--$4$ kHz, its gradient becomes steep (see the small panel
in Fig.~\ref{fig:fig2} for $f \agt f_{\rm cut}$). For $f \alt f_{\rm cut}$,
the merged object has a binary-like structure (i.e., there exist two
density maxima) which enhances emissivity of GWs, whereas at $f \sim
f_{\rm cut}$, such structure is disrupted, resulting in a quick
decrease of the GW amplitude.  A hump, which appears for $4 \alt f \alt 7$
kHz with the peak frequency $f_{\rm peak} \sim 5$--6 kHz, reflects the
formation and evolution of the spiral arms orbiting the central
object. For $f > f_{\rm peak}$, $h_{\rm eff}$ exponentially decreases;
this is the typical feature for the spectrum associated with a QNM
ring-down of the formed BH.  These spectrum features are qualitatively
universal as shown in Fig.~\ref{fig:fig2}. However, the 
spectrum shape depends quantitatively on the EOS, $m_0$, and $\nu$. 

The spectrum shape for $f \agt f_{\rm cut}$ is characterized by
$f_{\rm cut}$, $f_{\rm peak}$, and peak amplitude and width of the
hump ($h_{\rm peak}$ and $\sigma$), for which a schematic figure is
described in the small panel of Fig.~\ref{fig:fig2}.  We perform a
fitting procedure to extract these characteristic quantities from the
spectrum. As the first step, the spectrum is divided into two parts; one
is for $f \leq f_{\rm d}$, inspiral and early merger phases, and the
other is for $f\geq f_{\rm d}$, the merger and ringdown
phases. $f_{\rm d}$ is typically chosen as 4 kHz. For $f \leq f_{\rm
d}$, we fit the spectrum by $f^{-1/6}h_0/[1+{\rm exp}\{(f-f_{\rm
cut})/\Delta f\}]$. For $f\geq f_{\rm d}$, we adopt the Gaussian
distribution $h_{\rm peak}{\rm exp}[-(f-f_{\rm peak})^2/\sigma^2]$. In
the fitting, $f_{\rm cut}$, $\Delta f$, $h_0$, $h_{\rm peak}$, $f_{\rm
peak}$, and $\sigma$ are determined by the $\chi$-square fitting
\cite{Recipies}.


\begin{figure}
\centerline{
\includegraphics[width=7.8cm]{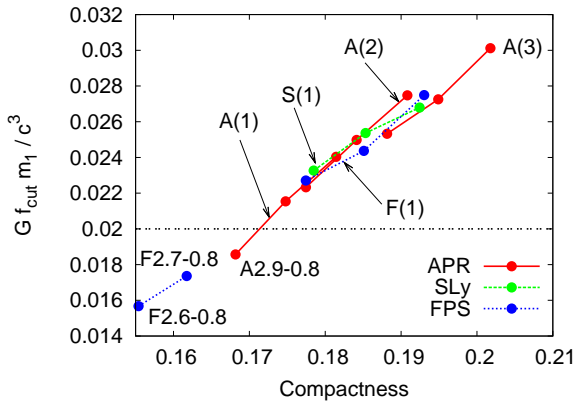}
}
\caption{\label{fig:fig3} ${\rm G}f_{\rm cut}m_1/{\rm c}^3$ as a
function of compactness of less massive binary component for all the
models.  Sequence A(1) represents the models A2.9-0.8, A3-0.8, and
A3.1-0.8 from left to right.  In a similar way, A(2) is a sequence of
A2.9-0.9, A3-0.9, and A3.1-0.9, A(3) is A2.9-1, A3-1, and A3.1-1, S(1)
is S2.8-1, S2.9-1, and S3-1, and F(1) is F2.6-1, F2.7-1, and
F2.8-1. The models below the horizontal line of ${\rm G}f_{\rm
cut}m_1/{\rm c}^3=0.02$ can produce disks of mass $\geq 0.01M_\odot$.
}
\end{figure} 

Figures~\ref{fig:fig3} plots ${\rm G}f_{\rm cut}m_1/{\rm c}^3$ as a
function of compactness of less massive companion defined by ${\rm
  G}m_1/c^2 R$ where $R$ is its circumferential radius.  ${\rm
  G}f_{\rm cut}m_1/c^3$ correlates strongly with ${\rm G}m_1/c^2 R$.
${\rm G}f_{\rm cut}m_1/c^3$ depends also on $m_0$; for the larger
value of $m_0$, it is larger for a given EOS and $\nu$.  Note that
larger values of $m_0$ and $\nu$ result in smaller values of disk mass
(see Fig. 4).  Thus, $f_{\rm cut}$ is an indicator that shows the
occurrence of the tidal disruption and possible disk formation.  (We
note that to obtain $f_{\rm cut}m_1$, we have to determine $m_1$ from
the gravitational-wave signal in the inspiral phase \cite{CF}. 
It is also worthy to note that $R$ may be determined, if
$f_{\rm cut}$ is measured and $m_1$ is determined; see
Ref.~\cite{Read:2009yp} for the related topic.)

\begin{figure}
\centerline{
\includegraphics[width=7.8cm]{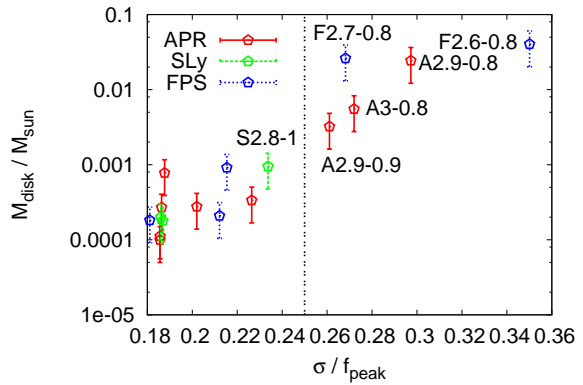}
}
\caption{\label{fig:fig4}
Disk mass as a function of $\sigma / f_{\rm peak}$ 
for all the models.}
\end{figure}

Figures~\ref{fig:fig4} plots the mass of disk surrounding a BH as a
function of $\sigma/f_{\rm peak}$. 
This shows that the disk mass
has a positive correlation with $\sigma/f_{\rm peak}$. 
On the other hand, we find that the disk mass does not have a clear
correlation with $h_{\rm peak}$.  The disk mass also correlates with
$m_0-M_{\rm thr}$ and $\nu$: $m_0-M_{\rm thr} \alt 0.2M_{\odot}$ and
$\nu \alt 0.8$ appear to be necessary for producing a massive disk
with $\geq 0.01M_\odot$ (see A2.9-0.8, F2.6-0.8, and F2.7-0.8).  The
disk mass for other models which do not satisfy this condition is
$10^{-3}$--$10^{-5}M_\odot$ \cite{Shibata:2002jb,Kiuchi:2009jt}.
Combination of Figs. 3 and 4 also proposes that a necessary condition
for producing a disk of mass $\agt 0.01M_{\odot}$ is ${\rm G}f_{\rm
cut} m_1/c^3\leq 0.02$ and $\sigma/f_{\rm peak} \agt 0.25$.


The frequency of GWs in the merger phase is slightly outside the most 
sensitive frequency band of the first and second generation detectors 
such as LIGO and advanced LIGO (10--1 kHz), because the values of
$f_{\rm cut}$ and $f_{\rm peak}$ are $\sim 2.5$--7 kHz.  Nevertheless,
the amplitude at $f=f_{\rm cut}$ and $f_{\rm peak}$ is not extremely
small for a hypothetical distance $D=100$ Mpc.  If the proposed third
generation detectors such as Einstein telescope become available with
a special design for high-frequency sensitivity, these quantities may
be measured. Our results indicate that the formation and evolution
processes of a BH and surrounding disk are imprinted in GW
spectrum. Also, GWs are the unique observable which directly carries
information of the central region of GRBs.  Thus, it is worthy to
explore methods for extracting physical information on the merger of
BNSs collapsing to a BH from observed GWs, which may clarify the
formation process of short GRBs.

\emph{Constraining the merger process by GW observation.}---As
mentioned in Introduction, the merger hypothesis for the central
engine of short GRBs requires that the mass of disk around a BH is
greater than $\sim 0.01M_\odot$. Our numerical results show that disk
mass correlates strongly with $\sigma/f_{\rm peak}$ and ${\rm G}f_{\rm
cut}m_1/c^3$ irrespective of EOS.  Assuming that GWs from BNSs are
frequently observed, and also, GWs and short GRB are observed
simultaneously for several events in the future, we propose a strategy
for exploring the merger hypothesis by analyzing GWs.  As a first
step, we should determine $m_0$, $\nu$, $f_{\rm cut}$, $f_{\rm
peak}$, and $\sigma$ for each event: The values of $m_0$ and $\nu$
will be determined from the inspiral waveform using the matched
filtering technique~\cite{CF}, and $f_{\rm cut}$, $f_{\rm
peak}$, and $\sigma$ be from the merger waveform.  As a second step,
we should infer the disk mass from ${\rm G}f_{\rm cut}m_1/c^3$ and
$\sigma/f_{\rm peak}$.  If the conditions of ${\rm G}f_{\rm
cut}m_1/c^3 \leq 0.02$ and $\sigma/f_{\rm peak} \agt 0.25$ are always
satisfied for events in which GWs and short GRB are simultaneously
observed, the merger hypothesis will be strongly supported.  Because a 
small value of ${\rm G}f_{\rm cut}m_1/c^3$ (for the typical mass of
neutron stars 1.3--$1.5M_{\odot}$) implies that mass ratio is much
smaller than unity, we could conclude that the origin of short GRBs
would be the merger of unequal-mass BNSs.  By contrast, if
simultaneous detections occur irrespective of the values of ${\rm
G}f_{\rm cut}m_1/c^3$ and $\sigma/f_{\rm peak}$, this does not agree
with our numerical results, because these should correlate with the
disk mass.  In such case, we have to conclude that an unknown
mechanism plays a role for producing the GRBs.

Our numerical results indicate that massive disk is not formed for
$\nu \approx 1$ or $m_0 > 3M_{\odot}$ for any EOS. This implies that
the merger of BNS collapsing to a BH does not always produce GRBs. 
For such case, only GWs will be observed. 
This suggests another test for the merger scenario.

Finally, a comment is given for the case that an HMNS is formed.
Because the HMNS should eventually collapse to a BH via gravitational
radiation \cite{Shibata:2002jb} or angular momentum transport by
magnetohydrodynamic effect \cite{DLSSS} or neutrino cooling, the HMNS
formation scenario can be an alternative in the merger hypothesis. In
the HMNS formation, the GW spectrum for the merger phase is
significantly different from that in the BH formation because
quasiperiodic GWs emitted by a quasiperiodic rotation of the HMNS is
likely to produce sharp peaks in the spectrum and characterize the
spectrum \cite{Shibata:2002jb}. In this case, detecting these peaks of
GWs will play a crucial role for confirming the merger hypothesis.



\emph{Acknowledgments.}--- Numerical computations were performed on
XT4 at the Center for Computational Astrophysics in NAOJ and on
NEC-SX8 at YITP in Kyoto University.  This work was supported by
Grant-in-Aid for Scientific Research (21340051) and for Scientific
Research on Innovative Area (20105004) of the Japanese MEXT, by
NSF Grant PHY-0503366, and by 
Grant-in-Aid of the Japanese Ministry of Education, Science, Culture,
 and Sport (21018008,21105511).



\end{document}